\RequirePackage{fix-cm}
\documentclass[a4paper,11pt]{llncs}       % onecolumn (second format)
\usepackage[utf8x]{inputenc}
\usepackage{hyperref}
\usepackage{graphicx,amsmath,amssymb}
\usepackage{mathtools}
\usepackage{forest}
\usepackage{comment}
\usepackage{tikz}
\usepackage{xspace}
\usepackage{standalone}
\usetikzlibrary{positioning, decorations.pathmorphing,arrows,chains}
\usetikzlibrary{decorations,decorations.text,calligraphy,arrows.meta}
\usetikzlibrary{decorations.pathmorphing,decorations.pathreplacing}
\usetikzlibrary{patterns}
\usepackage{pgfplots}
\usepackage{subcaption}
\usepackage{siunitx}
\usepackage{todonotes}
\usepackage{algorithm}
\usepackage[noend]{algorithmic}
\usepackage{multicol,multirow}
\usepackage{booktabs}
\usepackage[inline]{enumitem}

\usepackage{listings}
\usepackage{xcolor}

\newcommand{\NN}{\mathbb{N}}
\newcommand{\ZZ}{\mathbb{Z}}
\newcommand{\QQ}{\mathbb{Q}}
\newcommand{\RR}{\mathbb{R}}

\def\<#1>{\langle#1\rangle}
\DeclareMathOperator{\lc}{lc}
\DeclareMathOperator{\lm}{lm}

\DeclareMathOperator{\amba}{\mathfrak{a}}
\DeclareMathOperator{\SPol}{\textup{S-Pol}}

\newcommand{\name}{{\sf f4ncgb}}

\newcommand{\mathematica}{{\sf Mathematica}\xspace}
\newcommand{\ncalgebra}{{\sf NCAlgebra}\xspace}
\newcommand{\apcocoa}{{\sf ApCoCoA}\xspace}
\newcommand{\ncpoly}{{\sf NCPoly}\xspace}
\newcommand{\gbnp}{{\sf GBNP}\xspace}
\newcommand{\gap}{{\sf GAP}\xspace}
\newcommand{\operatorgb}{{\sf operator\_gb}\xspace}
\newcommand{\sagemath}{{\sf SageMath}\xspace}

\newcommand{\bergman}{{\sf bergman}\xspace}
\newcommand{\letterplace}{{\sf Letterplace}\xspace}
\newcommand{\ssingular}{{\sf Singular}\xspace}
\newcommand{\magma}{{\sf Magma}\xspace}
\newcommand{\macaulay}{{\sf Macaulay2}\xspace}
\newcommand{\opal}{{\sf Opal}\xspace}

\newcommand{\symbolicdata}{{\sf SymbolicData}\xspace}
\newcommand{\license}{MIT\xspace}

\begin{document}

\title{\name{}: High Performance Gr\"obner Basis Computations in Free
  Algebras\thanks{M.H.~was supported by the Austrian Science Fund
    (FWF) [10.55776/COE12].\\C.H.~was supported by the LIT AI Lab
    funded by the state of Upper Austria.} }

% Author name(s)
 \author{Maximilian Heisinger \and 
 	Clemens Hofstadler}

\institute{Institute for Symbolic Artificial Intelligence,
		Johannes Kepler University Linz,\\Linz, Austria\\
		\email{\{maximilian.heisinger,clemens.hofstadler\}@jku.at}
		}

\maketitle

{\setlength\emergencystretch{\hsize} % Fix spacing on the side of the abstract for this paragraph.
\begin{abstract}
We present \name{}, a new open-source C++ library for Gr\"obner basis computations in free algebras,
which transfers recent advancements in commutative Gr\"obner basis software to the noncommutative setting.  
As our experiments show, \name{} establishes a new state of the art for noncommutative Gr\"obner basis computations.
We also discuss implementation details and design choices.
\end{abstract}}


 \section{Introduction}

Over the last decades, Gr\"obner bases have become an indispensable tool in computational algebra and related fields.
Originally introduced by Bruno Buchberger in his seminal PhD thesis~\cite{Buc65},
Gr\"obner bases offer an algorithmic framework for solving a wide range of problems in polynomial algebra. 
While initially developed for commutative polynomials, the theory has since been extended to various noncommutative contexts, 
including free (associative) algebras~\cite{Ber78,Mor85}. 
In this setting, Gr\"obner bases (sometimes also referred to as {Gr\"obner-Shirshov bases}~\cite{Bok76}) 
have found important applications in (linear) control theory~\cite{HSW98}, automated theorem proving~\cite{HW94,CHRR20,SL20,Hof23}, 
as well as in graph~\cite{LESSSW22} and game theory~\cite{game-theory}.

All these applications crucially rely on the ability to efficiently compute Gr\"obner bases in free algebras.
While software for commutative Gr\"obner basis computations has seen remarkable progress in recent years, 
driven by the development of specialized data structures and optimized algorithms, noncommutative tools seem to lag behind. 
They often lack the same level of efficiency and sophistication, and mostly rely on outdated algorithms; see Section~\ref{sec:software} for a more detailed discussion.

In this work, we present \name{}, a new open-source C++ library for Gr\"obner basis computations in free algebras, distributed under the license \license
and available at \url{https://gitlab.sai.jku.at/f4ncgb/f4ncgb}.
The library supports noncommutative polynomial systems with coefficients in prime fields of characteristic $< 2^{31}$ and in the field of rational numbers.
To the best of our knowledge, \name{} is the first open-source tool to port the newest advancements in Gr\"obner basis computations from the commutative setting
to the noncommutative world.
Our experiments in Section~\ref{sec:experiments} show that \name{} establishes a new state of the art for noncommutative Gr\"obner basis computations among freely available tools.

The foundation of our implementation are specialized data structures for representing and storing noncommutative monomials and polynomials.
We abstract these data types into a \emph{store} structure, using a contiguous memory area for storing elements and a hash map for efficient lookup of known entries. 
We describe this structure in more detail in~Section~\ref{sec:ds}.

Algorithmically, the core of \name{} is an efficient implementation of Faugère's F4 algorithm. 
Originally developed for commutative polynomials~\cite{Fau99} and later adapted to free algebras~\cite{kang2007linear,Xiu12}, 
the F4 algorithm reduces Gr\"obner basis computations to normal form computations of large sparse matrices via Gaussian elimination. 

To efficiently support this, we have implemented a specialized Gaussian elimination algorithm tailored for sparse matrices, 
based on ideas from the commutative case~\cite{MP15}.
Our implementation makes use of modern multi-core hardware by parallelizing matrix reductions, resulting in significant overall speedups. 
For inputs with rational coefficients, we employ a multi-modular strategy: computations are carried out modulo different primes, 
followed by a reconstruction step to recover the rational solution.
We provide an overview of the noncommutative F4 algorithm in Section~\ref{sec:f4} and discuss our linear algebra implementation in~Section~\ref{sec:lin-alg}.

In addition to computing Gr\"obner bases, \name{} offers proof logging capabilities in the form of \emph{cofactor representations}:
for each polynomial in the output basis, a representation in terms of the input can be computed, see Section~\ref{sec:proofs}.
These representations are not only of interest for verifying the correctness of the output, but are also useful in applications~\cite{HRR19}.

\section{Related Software}
\label{sec:software}

There already exists a wide range of tools for Gr\"obner basis computations in free algebras.
In this section, we provide an overview of related software packages documented in the literature.

While the landscape of tools is diverse, almost all available packages share a common problem: they lag behind in terms of algorithmic development. 
In the commutative setting, essentially all modern Gr\"obner basis implementations rely on Faugère's F4 algorithm, 
which provides substantial performance improvements over earlier methods.
By contrast, almost all noncommutative packages continue to use the older and generally less efficient Buchberger algorithm.

Notable examples of tools relying on Buchberger's algorithm 
include the \mathematica library \ncalgebra~\cite{ncalgebra}, the \apcocoa package \ncpoly~\cite{ncpoly},
and the \gap package \gbnp~\cite{gbnp}.
Other tools that also fall into this category are the Standard Lisp implementation \bergman~\cite{bergman} and the C++ tool \opal~\cite{opal}, which both 
seem no longer actively maintained.

An important exception is the subscription-based computer algebra system \magma~\cite{magma}, 
which is one of the few tools to implement a noncommutative version of the F4 algorithm. 
In addition, \magma also features dedicated data structures and highly optimized subroutines (e.g., for the linear algebra),
making it arguably the state of the art for computing noncommutative Gr\"obner bases. 
However, a significant drawback is its proprietary nature, which limits its availability to the community.

For homogeneous ideals over prime fields, the open-source system \macaulay~\cite{M2} offers a decent implementation of the F4 algorithm.
For all other ideals, however, it falls back to Buchberger's algorithm.

To the best of our knowledge, the only open-source implementation of the noncommutative F4 algorithm over the field of rational numbers is the \sagemath package \operatorgb~\cite{operatorgb}, developed by one of the authors.
While \operatorgb already implements the noncommutative F4 algorithm, it still relies on relatively simple data structures for representing and manipulating noncommutative objects.
Additionally, the package is primarily written in interpreted Python, with selected components (such as the linear algebra) in compiled Cython, which limits performance.
As a result, it cannot match the efficiency of the highly optimized and fully compiled C code in \magma.
The new package \name{} can be viewed as a significant advancement of \operatorgb, offering both improved algorithmic design and a more efficient implementation.

Special mention should also be given to the \letterplace~\cite{letterplace} subsystem of \ssingular, 
which has established itself as the standard for freely available noncommutative Gr\"obner basis software.
It offers a good tradeoff between efficiency and functionality, 
supporting a range of methods that go well beyond Gr\"obner basis computation alone (see~\cite{letterplace} for an overview).
\letterplace is based on the \emph{letterplace correspondence}~\cite{letterplace-old}, 
a technique that embeds noncommutative computations into a larger commutative polynomial ring.
This embedding allows for the reuse of optimized commutative data structures and subroutines in the noncommutative setting.

We see two notable limitations in the letterplace approach.
The first one is inherent to the letterplace correspondence itself: the number of variables in the commutative ring grows linearly with the degree
of the considered noncommutative polynomials.
As a result, for high-degree inputs, the commutative computation can become prohibitively large.
The second limitation is an implementation issue of \ssingular, as it still relies on Buchberger's algorithm even for commutative computations.

\section{Preliminaries -- Noncommutative Gr\"obner Bases}
\label{sec:prelim}

To keep this work self-contained, we recall the most important aspects of the Gr\"obner basis theory in free algebras. 
For further details including proofs, we refer interested readers to, e.g.,~\cite{Xiu12,Mor16,Hof23}.

We write $\< X>$ for the \emph{free monoid} on a finite set of indeterminates $X = \{x_1,\ldots,x_n \}$.
Recall that $\<X>$ consists of all finite words over the alphabet $X$, including the empty word $1$.
Multiplication in $\<X>$ is given by concatenation of words, 
which also induces a natural notion of divisibility: for $v,w \in \<X>$, we say that $v$ divides~$w$ if $v$ is a subword of~$w$, i.e.,
if there exist $a,b \in \<X>$ such that $w = avb$. 

For a field $K$, we consider the \emph{free (associative) algebra} $K\<X>$.
Elements in $K\<X>$ are finite sums of the form $c_1 m_1 + \cdots + c_d m_d$ with nonzero $c_1,\dots,c_d \in K$ and pairwise different $m_1,\dots,m_d \in \<X>$.
We consider these elements as \emph{(noncommutative) polynomials} with coefficients in $K$ and monomials in $\<X>$.

For a set of polynomials $F \subseteq K\<X>$, we let $(F)$ be the (two\nobreakdash-sided) ideal generated by $F$, 
which consists of all two-sided linear combinations of the elements in $F$ with polynomials as coefficients, i.e.,
$(F) = \left\{\sum_{i=1}^k p_i f_i q_i \mid f_i\in F,\; p_i, q_i\in K\<X>,\; k\in\NN\right\}$.
A set $F$ is a \emph{generating set} of an ideal $I \subseteq K\<X>$ if $I = (F)$.
Recall that if $|X| > 1$, then there exist ideals in $K\<X>$ that do not possess a finite generating set. 
A classical example is the ideal $( \{ xy^n x \mid n \in \NN \}) \subseteq K\<x,y>$.

A total ordering $\prec$ on $\left\langle X\right\rangle$ is called a 
\emph{monomial (well-)ordering} if the following conditions hold:
\begin{enumerate}
        \item $\forall a,b,v,w \in \left\langle X\right\rangle$: if $v\prec w$, then $avb \prec awb$;
        \item every nonempty subset of $\<X>$ has a least element.\label{cond:monomial-order-2}
\end{enumerate}

A very general class of monomial orderings are \emph{weighted block orderings} of the following form:
Fix $d \in \NN_{> 0}$ and assign to each variable $x_{i} \in X$ a nonzero weight vector $w_{i} \in \RR_{\geq 0}^{d}$. 
Define the weight $w(m)$ of a monomial $m = x_{i_{1}}\dots x_{i_{k}} \in \<X>$ as $w(m) = \sum_{j=1}^{k} w_{i_{j}} \in \RR_{\geq 0}^{d}$.
The weighted block ordering $\prec_{\omega}$ w.r.t.~the weight matrix $\omega = (w_{1},\dots,w_{n})$ is given by:
$$
	m \prec_{\omega} m' \quad \text{iff}\quad w(m) < w(m') \text{ or } \left(w(m) = w(m') \text{ and } m \prec_{\textup{lex}} m'\right),
$$
where the real-valued vectors $w(m)$ and $w(m')$ are compared lexicographically and
$m \prec_{\textup{lex}} m'$ denotes a lexicographic comparison of the words $m$ and $m'$ from left to right.

At the moment, our implementation supports weighted block orderings where each weight vector is a unit vector $(0,\dots,0,1,0,\dots,0)$.
This is sufficient to represent classical orderings such as degree-lexicographic orderings.
It also enables the use of elimination orderings~\cite[Def.~2.4.15]{Hof23}, 
which are crucial, e.g., for computing elimination ideals~\cite[Thm.~2.4.43]{Hof23}.

For a fixed monomial ordering, the \emph{leading monomial} $\lm(f)$ and \emph{leading coefficient} $\lc(f)$ of a nonzero polynomial $f\in K\<X>$ are defined in the usual way.
In our definition, the leading monomial does not contain a coefficient.

\begin{definition}
A \emph{Gr\"obner basis} of an ideal $I \subseteq K\<X>$ (w.r.t.~a monomial ordering $\prec$) is a subset  
$G\subseteq I$ such that for every nonzero $f \in I$, there exists $g \in G$ such that $\lm(g)$ divides $\lm(f)$.
\end{definition} 

We note that all standard characterizations of Gr\"obner bases from the commutative setting carry over naturally to the noncommutative case; see, e.g.,~\cite[Thm.~2.4.37]{Hof23}.
In particular, Buchberger’s criterion~\cite[Thm.~2.6.6]{cox-little-oshea}, 
which provides an effective test for whether a set of commutative polynomials forms a Gr\"obner basis, 
has a noncommutative analogue in \emph{Bergman’s diamond lemma}~\cite[Thm.~1.2]{Ber78}.

To state this result, we recall the concept of \emph{ambiguities} as introduced in~\cite{Ber78}.
In the following, we use $\lvert w \rvert$ to denote the length of a word~$w \in \<X>$.

\begin{definition}\label{def:ambiguities-normal}
Let $f,g \in K\<X> \setminus \{0\}$ and let $a,b,c,d \in \<X>$ be such that one of the following conditions holds:
\begin{alignat*}{3}
	&\textit{1.}~\lm(af) &&=\, \lm(gd) &&\text{ and $b = c = 1$ and $1 \leq \lvert a \rvert < \lvert \lm(g) \rvert$},\\
	&\textit{2.}~\lm(fb) &&=\, \lm(cg)  &&\text{ and $a = d = 1$ and $1 \leq \lvert b \rvert < \lvert \lm(g) \rvert$},\\
	&\textit{3.}~\lm(afb) &&=\, \lm(g) &&\text{ and $c = d = 1$ and $f \neq g$},\\
	&\textit{4.}~\lm(f) &&=\, \lm(cgd)\; &&\text{ and $a = b = 1$ and $f \neq g$}.
\end{alignat*}  
Then we call the tuple $(a\otimes b, c\otimes d, f, g)$ an \emph{ambiguity} of $f$ and $g$.
\end{definition}

Note that two elements can give rise to multiple ambiguities, but only to finitely many. 
A polynomial may also form ambiguities (of type 1 and~2) with itself.
If $\amba = (a\otimes b, c \otimes d, f, g)$ is an ambiguity, then $\lm(a f b) = \lm(c g d)$ by construction.
Based on this, we define the \emph{degree} of $\amba$ to be $\deg(\amba) = \lvert \lm(afb) \rvert = \lvert \lm(cgd) \rvert$.

Analogous to Buchberger’s criterion, Bergman’s diamond lemma is based on a notion of (noncommutative) polynomial reduction,
which is (basically) identical to the reduction used in the commutative case: $f$ can be reduced by $g$ if $\lm(g)$ divides a monomial in $f$ and the result of the reduction is $f - \lambda vgw$ with $\lambda \in K$, $v,w \in \<X>$ chosen so that the targeted monomial cancels. 
For a formal definition, see, e.g.,~\cite[Sec.~2.4.2]{Hof23}.

An ambiguity of two polynomials $f$ and $g$ reflects the fact that the monomial $\lm(afb) = \lm(cgd)$ can be reduced in two (possibly different) ways by $f$ and $g$.
A set $G$ containing $f$ and $g$ can only be a Gr\"obner basis if these two reductions lead to the same canonical normal form.
To test whether this is the case, one considers the associated \emph{S-polynomial}.

\begin{definition}\label{def:S-polynomial}
Let $G \subseteq K\<X>$ and let $\amba = (a\otimes b, c \otimes d,f,g)$ be an ambiguity of $f,g\in G$.
The \emph{S-polynomial} of $\amba$ is
\[
	\SPol(\amba) = \frac{1}{\lc(f)} a f b - \frac{1}{\lc(g)} c g d.
\]
\end{definition}

\begin{theorem}[Bergman's diamond lemma]
\label{thm:diamond-lemma}
A generating set $G$ of an ideal $I \subseteq K\<X>$ is a Gr\"obner basis of $I$ 
if and only if, for all ambiguities $\amba$ of $G$, the S-polynomial $\SPol(\amba)$ reduces to $0$ by $G$.
\end{theorem}

Theorem~\ref{thm:diamond-lemma} provides an effective criterion to test whether a given set forms a Gr\"obner basis.
It also directly gives rise to a completion procedure for constructing a Gr\"obner basis from an arbitrary generating set --
a noncommutative analogue of Buchberger's algorithm; see~\cite[Alg.~1]{Hof23} for a detailed description.

However, in contrast to the commutative case, not all finitely generated ideals in the free algebra admit a finite Gr\"obner basis (w.r.t.~a fixed monomial ordering).
For instance, the principal ideal $(xyx - xy) \subseteq \QQ\<x,y>$ does not admit a finite Gr\"obner basis for any choice of ordering.
  
As a consequence, we cannot expect such a completion algorithm to terminate in general.
Instead, we have to content ourselves with an enumeration procedure with the following behaviour:
\begin{enumerate}
	\item If the input ideal $I$ admits a finite Gr\"obner basis (w.r.t.~the chosen monomial order), the procedure terminates and returns such a basis.
	\item Otherwise, it enumerates an infinite sequence $(g_n)_{n \in \NN}$ such that the set $\{g_n \mid n \in \NN\}$ forms a Gr\"obner basis of $I$.
\end{enumerate}

\section{Noncommutative F4 Algorithm}
\label{sec:f4}

Our library \name{} implements the noncommutative F4 algorithm for computing Gr\"obner bases in free algebras.
Below, we provide a brief overview of the algorithm; for a detailed exposition, see~\cite[Sec.~5.4]{Xiu12}. 
For those familiar with the commutative algorithm, we note that the noncommutative version follows the same high-level structure.

After initializing the prospective Gr\"obner basis with the input polynomials and computing all ambiguities between them, 
the F4 algorithm enters a loop consisting of the following main steps:
\begin{enumerate}
	\item \textbf{Critical pair selection}: Select a subset of ambiguities, usually all those of minimal degree.
	For each selected ambiguity $\amba = (a\otimes b, c\otimes d, f, g)$, form the polynomials $afb$ and $cgd$.
	\item \textbf{Symbolic preprocessing}: Iterate over all monomials appearing in the polynomials generated in the
	previous step.\label{item:f4-sym-pre}
	For each monomial~$m$, check whether it is divisible by the leading monomial of a basis element~$g$.
	If so, form a polynomial $vgw$ with $v,w \in \< X >$ such that $\lm(vgw) = m$.
	If multiple choices for $v,g,w$ are possible, pick one (for~$g$, e.g., the one with minimal leading monomial).
	Repeat this process also for all monomials in the newly created polynomials.
	\item  \textbf{Matrix construction}: Collect all polynomials constructed in the previous two steps in a matrix.
	To this end, fix an arbitrary order of the polynomials $f_1, f_2, \dots$ 
	and order all monomials appearing in those polynomials in descending order $m_1 \succ m_2 \succ \dots$.
	The matrix entry at position $(i,j)$ is then given by the coefficient of monomial $m_{j}$ in polynomial $f_{i}$.
	\item  \textbf{Matrix reduction}: Reduce the matrix constructed in the previous step to 
	(reduced)\footnote{In principle, a row echelon form suffices. However, typically using the 
	reduced row echelon form improves the algorithm’s overall performance.}
	row echelon form.\label{item:f4-reduction}
	\item \textbf{Basis update}: Translate the rows of the reduced matrix back into polynomials
	and add all those polynomials to the basis whose leading monomial is not divisible by an element of the current basis.
	Then, form all new ambiguities.
	Optionally, one can apply the noncommutative variant of the Gebauer-M\"oller criteria~\cite[Thm.~4.2.22]{Xiu12} to discard redundant ambiguities.
	If ambiguities remain, continue with the next iteration.\label{item:f4-update}
\end{enumerate} 

This procedure enumerates a Gr\"obner basis in the sense of the previous section. 
To guarantee termination in practice, we have to add additional constraints.
One way of doing this is by simply stopping the computation after a prescribed number of iterations.
Another way is to impose an upper bound on the degree of ambiguities that are considered.
If, during the execution of the algorithm, an ambiguity arises whose degree is larger than the designated bound, this ambiguity is discarded.
Both approaches ensure termination but only yield a partial Gr\"obner basis.
Our implementation supports both termination options.

\section{Basic Data Structures -- Monomials and Polynomials}
\label{sec:ds}

Our library is based on specialized
data structures for handling and storing noncommutative monomials and polynomials. 
%We present their design and tradeoffs in this section.
The core operations that these data structures have to support are:
\begin{enumerate}
\item For monomials: comparison w.r.t.~monomial ordering, multiplication, and divisibility tests. 
\item For polynomials: leading monomial extraction and multiplication by monomials.
\end{enumerate}

In our implementation, we represent each variable by a positive integer, 
which are assigned in increasing order according to the given monomial ordering, i.e., the smallest variable is assigned $1$, 
the next smallest $2$, and so on.
Each noncommutative monomial then corresponds to a unique sequence of integers, with the empty sequence denoting the empty monomial $1$. 
This encoding allows lexicographic comparison between monomials to reduce to a fast lexicographic comparison of integer sequences.

%Our implementation of both monomials and polynomials share some common properties. They are
%\begin{enumerate*}[label=(\roman*)]
%\item unique,
%\item immutable,
%\item of constant-size, i.e., their element count does not change after they are instantiated, and
%\item read very often, with a high chance of re-use.
%\end{enumerate*}

To store monomials (and also polynomials), we have designed a common base
\emph{store} structure that is extended into \emph{monomial store} and
\emph{polynomial store} structures using the \emph{Curiously Recurring
  Template Pattern}~\cite{10.5555/260627.260647}. The store implements
a bump-pointer allocator combined with an open-addressing hash map (using
Boost's \texttt{unordered\_flat\_map}) for efficient lookups of known entries.

Each element in a store consists of some associated metadata followed by the data of the element itself.
The metadata contains the length of the element (for a polynomial this is the number of terms) and optionally additional data.

\begin{figure}[t]
   \centering
   \begin{tikzpicture}

% Example: 2x+5xy+xyx-xyz
\node[align=center, anchor=south,rotate=90] at (0,0.5) {{\scriptsize Monomials}};
  
% Entry 0: x
\draw[thick,fill=gray!30] (0,0) -- (0,1) -- (1,1) -- (1,0);
\node [align=center] at (0.5,0.5) {len\\1};
\draw (0,1) node[anchor=south west] {0} -- (0,1.3);
\node [align=center] at (1.25,0.5) {1};
\draw[thick,pattern=north west lines, pattern color=gray!70] (1.5,0) -- (1.5,1) -- (2,1) -- (2,0);

% Entry 1: xy
\draw[thick,fill=gray!30] (2,0) -- (2,1) -- (3,1) -- (3,0);
\node [align=center] at (2.5,0.5) {len\\2};
\draw (2,1) node[anchor=south west] {8} -- (2,1.3);
\node at (3.25,0.5) {2};
\node at (3.75,0.5) {1};

% Entry 2: xyx
\draw[thick,fill=gray!30] (4,0) -- (4,1) -- (5,1) -- (5,0);
\node [align=center] at (4.5,0.5) {len\\3};
\draw (4,1) node[anchor=south west] {16} -- (4,1.3);
\node at (5.25,0.5) {1};
\node at (5.75,0.5) {2};
\node at (6.25,0.5) {1};
\draw[thick,pattern=north west lines, pattern color=gray!70] (6.5,0) -- (6.5,1) -- (7,1) -- (7,0);

% Entry 3: xyz
\draw[thick,fill=gray!30] (7,0) -- (7,1) -- (8,1) -- (8,0);
\node [align=center] at (7.5,0.5) {len\\3};
\draw (7,1) node[anchor=south west] {28} -- (7,1.3);
\node at (8.25,0.5) {1};
\node at (8.75,0.5) {2};
\node at (9.25,0.5) {3};
\draw[thick,pattern=north west lines, pattern color=gray!70] (9.5,0) -- (9.5,1) -- (10,1) -- (10,0);

% Draw the main data stream
\draw[thick,decorate,decoration=lineto] (0,0) -- (10.5,0);
\draw[thick,decorate,decoration=lineto] (0,1) -- (10.5,1);
\draw[thick,decorate,decoration=zigzag,segment length=2mm] (10.5,1) -- (10.5,0);

% Add tick marks for 2-byte boundaries in the wide array
\foreach \x in {0,0.5,...,10}
    \draw (\x,0) node [anchor=north west, distance=0, inner sep=0.5mm] {\tiny \pgfmathparse{\x*4}\pgfmathprintnumber[fixed, precision=0]{\pgfmathresult}} -- (\x,-0.1);

\draw [decorate,decoration={brace,amplitude=5pt,raise=0.4cm}]
  (0,1) -- (1,1) node[midway,yshift=0.88cm]{metadata};

\draw [decorate,decoration={brace,amplitude=5pt,raise=0.4cm}]
  (6.5,1) -- (7,1) node[midway,yshift=0.88cm]{padding (for 32 bits)};

\draw [decorate,decoration={brace,amplitude=5pt,raise=0.4cm}]
  (3,1) -- (4,1) node[midway,yshift=0.88cm]{variables};

%%%%%%%%%%%%%%%%%%%%%%%%%%%%%%%%%%%%%%%%%%%%%%%%%%%%%%
% Polynomial
%%%%%%%%%%%%%%%%%%%%%%%%%%%%%%%%%%%%%%%%%%%%%%%%%%%%%%
  
\node[align=center, anchor=south,rotate=90] at (0,-1.5) {{\scriptsize Polynomials}};
\draw (0,-1) node[anchor=south west] {0} -- (0,-.7);
\draw[thick,fill=gray!30] (0,-2) rectangle ++(2,1);
\draw[thick] ({1},-2) -- ({1},-1);
\draw[thick] ({2},-2) -- ({2},-1);

\node [align=center] at (0.5,-1.5) {len\\4};
\node [align=center] at (1.5,-1.5) {coeffs\\\mbox{}};

\filldraw (1.5,-1.7) circle[radius=2pt]; % Solid dot
  \draw[-{Latex}, thick] (1.5,-1.7) -- (0.5,-3); % Arrow tip at the end

\node at (2.5,-1.5) {28};
\draw[thin] ({3},-2) -- ({3},-1);
\node at (3.5,-1.5) {16};
\draw[thin] ({4},-2) -- ({4},-1);
\node at (4.5,-1.5) {8};
\draw[thin] ({5},-2) -- ({5},-1);
\node at (5.5,-1.5) {0};
\draw[thick] ({6},-2) -- ({6},-1);

\foreach \x in {0,0.5,...,6}
    \draw (\x,-2) node [anchor=north west, distance=0, inner sep=0.5mm] {\tiny \pgfmathparse{\x*4}\pgfmathprintnumber[fixed, precision=0]{\pgfmathresult}} -- (\x,-2.1);

\draw[thick,decorate,decoration=lineto] (0,-2) -- (6.5,-2);
\draw[thick,decorate,decoration=lineto] (0,-1) -- (6.5,-1);
\draw[thick,decorate,decoration=zigzag,segment length=2mm] (6.5,-1) -- (6.5,-2);

%%%%%%%%%%%%%%%%%%%%%%%%%%%%%%%%%%%%%%%%%%%%%%%%%%%%%%
% Coefficients
%%%%%%%%%%%%%%%%%%%%%%%%%%%%%%%%%%%%%%%%%%%%%%%%%%%%%%

% Label on the left
%\draw (0,-3) node[anchor=south west] {0} -- (0,-2.7);
\node[align=center, anchor=south,rotate=90] at (0,-3.5) {{\scriptsize Coefficients}};

% Draw main gray box for 4 coefficients
\draw[thick] (0,-4) rectangle ++(4,1);

% Subdivide for each coefficient (c0, c1, c2, c3)
\foreach \x in {1,2,3,4}
    \draw[thin] (\x,-4) -- (\x,-3);

% Labels for coefficient IDs
\node at (0.5,-3.5) {$\frac{1}{2}$};
\node at (1.5,-3.5) {$2$};
\node at (2.5,-3.5) {$-1$};
\node at (3.5,-3.5) {$\frac{2}{3}$};

% Outline for top/bottom and right side (for visual consistency)
\draw[thick,decorate,decoration=lineto] (0,-4) -- (4.5,-4);
\draw[thick,decorate,decoration=lineto] (0,-3) -- (4.5,-3);
\draw[thick,decorate,decoration=zigzag,segment length=2mm] (4.5,-3) -- (4.5,-4);
\end{tikzpicture}
  \caption{
  Illustration of a monomial and polynomial store with global coefficient array over $\QQ\< x, y, z >$, using a monomial ordering with $x \prec y \prec z$. 
  The monomial store contains $x$, $yx$, $xyx$, and $xyz$ (in this order). 
  The polynomial store contains the polynomial $\frac{1}{2}xyz + 2xyx - yx + \frac{2}{3}x$, referencing monomials by index.
  The indices above the monomial and polynomial store are used throughout our program to reference the objects.}
  \label{fig:store}
\end{figure}
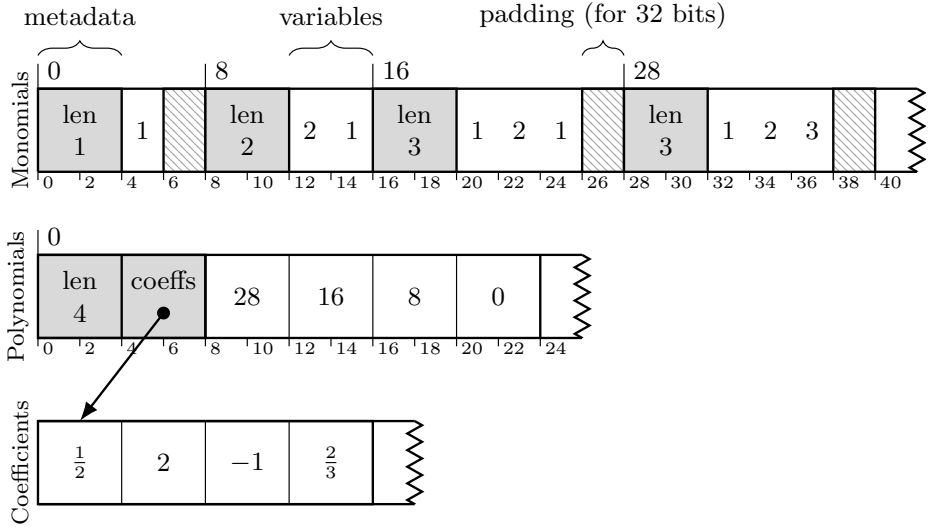

Both the metadata and the stored values of each element are correctly
aligned according to the requirements of their contained primitive
data types, positionally allocated and deallocated, and automatically
padded. We visualize the store's structure in Figure~\ref{fig:store}.

We rely on \texttt{malloc} to return \emph{uninitialized memory} in
order to allocate \SI{4}{\giga\byte} of sequential heap space when a
store is initialized. This uninitialized memory is not yet hardwired
into our process, instead its memory range is only reserved in our
process space, still respecting memory limits. New entries are
appended at the tip of the store, filling the allocated heap,
initializing the memory, and actually wiring it into our process
pagewise. This approach ensures that each appended element has a
constant address, making pointers to it stable. We can therefore use
pointers to elements during the entire program runtime.

%then does not need to store any other information on an element than a
%pointer, a length, and its accompanying index in the store. Another
%benefit of this structure is the increased memory locality, which
%manifests in more cache hits and increased performance, especially
%with the monomial store.

For a Gr\"obner basis computation, we maintain one global monomial store that contains all monomials generated during the computation.
Each monomial is uniquely identified by its index in this store, allowing for efficient referencing. 
When a new monomial is encountered, we first query the hash map to check whether it already exists in the store. 
If it does, we retrieve its index; if not, the sequence of integers representing the monomial is inserted into the store, along with its metadata, and assigned a new index.

Similarly, we also maintain a global polynomial store that contains all polynomials generated during a Gr\"obner basis computation.
Each polynomial entry consists of metadata and of a list of its monomials.
The monomials are represented by their indices in the monomial store and are ordered in descending order according to the monomial ordering, 
making the leading monomial easily accessible as the first entry. 
The metadata consists of the polynomial's length and of an index into a global coefficient array that stores the coefficients of all polynomials.
We do not store the coefficients directly in the polynomial entries, 
since typically many polynomials share the same coefficients (because they are monomial multiplies of each other). 
This approach allows us to store the coefficients only once and reference them from multiple polynomials, resulting in memory savings of up to \SI{40}{\percent}.
All rational coefficients are stored with arbitrary-precision.
Figure~\ref{fig:store} shows an example of a monomial and polynomial store.

Our store data structure allows for efficient storing and referencing of monomials (and polynomials), but it does not support fast divisibility tests 
-- an operation frequently required in the F4 algorithm.
In particular, monomial divisibility tests occur during symbolic preprocessing (step~\ref{item:f4-sym-pre}) 
and when computing ambiguities (step~\ref{item:f4-update}).
In both cases, we have to test whether a monomial is divisible by one of the leading monomials of the current Gr\"obner basis.

To handle this efficiently, we use a separate data structure: a \emph{prefix tree} (or \emph{trie}).
In such a tree, each node represents a variable, and each path from the root to a leaf encodes a monomial. 
When inserting a monomial, we traverse or construct the corresponding path based on its sequence of variables. 
Since common prefixes are shared, a prefix tree naturally supports fast prefix-based queries such as divisibility tests.

We maintain a global prefix tree consisting of all leading monomials of the current Gr\"obner basis, which is used for all divisibility tests.
As elements are never removed from the basis -- and hence from the prefix tree -- the tree can be stored in contiguous memory, 
improving cache locality and enabling faster traversals.

\subsection{Further Optimizations}
\label{sec:further-optimizations}

\paragraph{Scratch Space for Products}

In order for new elements to be looked up in a store's hash map, they
have to be constructed. Instead of using a separate buffer and a copy
operation for this, we introduce a \emph{scratch space} at the tip
of a store's heap. Each time a product of multiple monomials is
computed, the resulting monomial is constructed in that space. After
constructing the element, the store checks whether the element is already
contained using its hash map. If it is, the scratch is discarded and
later overwritten. If it is not, we \emph{commit} the scratch space,
storing the newly created element. We also update the hash map
accordingly, increase the element count, and move the tip of the store
to after the newly inserted element, including potential padding.

\paragraph{Memory Allocator and Tunables}
\sloppy
If available, we use \emph{mimalloc}~\cite{mimalloc} instead
of the system's default memory allocator. In many tests, this change
alone leads to a \SI{15}{\percent} speedup compared to the default. If
relying on \emph{glibc}, we recommend running \name{} with the
environment variable \texttt{GLIBC\_TUNABLES=glibc.malloc.hugetlb=1}.
We attribute these differences to a faster allocation and deallocation
speed (especially for hash map buckets) and to improved
coefficient storage.

%\paragraph{Purposefully Leaking Memory}
%
%Both the monomial and polynomial store structures are allocated once
%during the program's lifetime. They are deallocated at program exit,
%also deallocating every single coefficient with stored within the
%polynomial stores. Deallocation takes a long time for this, especially
%when using \emph{glibc}. We reduce the deallocation overhead by
%leaking the pointer to both stores, effectively delegating
%deallocation to the operating system. We automatically only enable
%this optimization on release builds without address sanitization
%enabled, as to not produce spurious warnings.

\subsection{Detrimental Optimizations}
\label{sec:detrimental-optimizations}

\paragraph{Saving Known Monomial Products}

Instead of the scratch space, we also tried to cache the results of monomial multiplications in a product hash map. 
However, this approach required more memory and resulted in a slowdown compared to using the scratch space alone.
While still surprising, this result can be explained by the reduced memory usage and improved cache locality of the scratch space method.
%We still offer this functionality through the \cmake{}-option \texttt{ENABLE\_MONOMIAL\_PRODUCTS\_MAP}.

\paragraph{Compact Store Membership}

To check whether an element already exists in the store, we use the store's hash map.
We tried to replace that hash map by a specialized hasher. 
Instead of hashing the elements directly, we store only the ID of each element from the store in a set.
To test whether a new element in the scratch space shall be committed to the store, we then check for set 
membership in this set, testing not for equality of the IDs, but instead of the data they point to. 
This, however, dramatically increased the time for store membership tests in our experiments. 
%We offer this feature behind the \cmake{}-flag \texttt{ENABLE\_COMPACT\_STORE\_MAP}.

\section{Linear Algebra}
\label{sec:lin-alg}

Computationally, the core operation of the F4 algorithm lies in the reduction of huge, structured sparse matrices to (reduced) row echelon form (step~\ref{item:f4-reduction}).
This step dominates the overall runtime and thus demands a highly optimized implementation.
In \name{}, we have implemented a specialized and parallelized Gaussian elimination procedure that exploits the structure of the matrices arising in Gr\"obner basis computations.

For computations over prime fields, our implementation closely follows the algorithm introduced in~\cite{MP15}. 
In the case of rational coefficients, we adopt a multi-modular approach, 
reducing rational computations to computations modulo different primes followed by a reconstruction step.

Since matrices arising in Gr\"obner basis computations typically only contain a few nonzero entries per row, 
we store them in a sparse row-wise format as an array of pointers to rows.
Each row is itself composed of two arrays: one containing the column indices of the nonzero entries (sorted in ascending order), and the other containing the corresponding values.

\subsection{Gaussian Elimination in Prime Fields}
\label{sec:gauss-Zp}

Our implementation supports prime fields with characteristic $< 2^{31}$.
This allows us to represent matrix entries using 32-bit unsigned integers, which ensures both memory efficiency and fast arithmetic operations.

To compute the reduced row echelon form of an $m \times n$ matrix $A$ over $\ZZ_{p}$ for a prime $p < 2^{31}$, we proceed in two steps: 
First, we transform $A$ into row echelon form, and then apply backward elimination to obtain the \emph{reduced} row echelon form.

\begin{figure}[tb]
  \centering
  \begin{tikzpicture}
  \def\n{15}  % number of rows
  \def\m{20}  % number of columns
  \def\cellsize{0.2} % cell size
  \def\nonzeroStart{{18,16,15,13,11,10,8,8,7,6,6,4,3,2,0}}

  \foreach \i in {0,...,13,7} {
    \pgfmathsetmacro{\start}{\nonzeroStart[\i]}
    \pgfmathsetmacro{\x}{\start * \cellsize}
    \pgfmathsetmacro{\y}{-\i * \cellsize}
    \pgfmathsetmacro{\width}{(\m - \start) * \cellsize}    
    \ifnum\i<7
      \fill[gray!70] (\x, \y) rectangle ++(\width, -\cellsize);
    \else\ifnum\i=7
      \foreach \j in {2,...,19} {
        \pgfmathsetmacro{\xj}{\j * \cellsize}
        \ifnum\j<8
        		\draw[fill=white] (\xj, \y) rectangle ++(\cellsize, -\cellsize);
	\else
		\draw[fill=gray!80] (\xj, \y) rectangle ++(\cellsize, -\cellsize);
	\fi
      }
      \draw[] (\x, \y) rectangle ++(\width, -\cellsize);
    \else
      \fill[gray!70] (\x, \y) rectangle ++(\width, -\cellsize);
    \fi\fi
  }
  \draw[->, thick] (-1, -1.5) -- (0.3, -1.5);
  \node[anchor=west,align=center,] at (-4, -1.4) {\footnotesize currently processed row\\[-0.2em] \footnotesize(dense buffer)};
  \draw[decorate, decoration={brace, mirror, amplitude=8pt}] 
    (20*\cellsize + 0.1, -7*\cellsize) -- (20*\cellsize + 0.1, 0) node[midway,xshift=10pt, right, font=\footnotesize,align=left] {already known\\pivot rows (sparse)};
\end{tikzpicture}
  \caption{Visualization of our Gaussian elimination over $\ZZ_{p}$. 
  Potentially nonzero entries are shaded in grey.}
  \label{fig:gauss}
\end{figure}
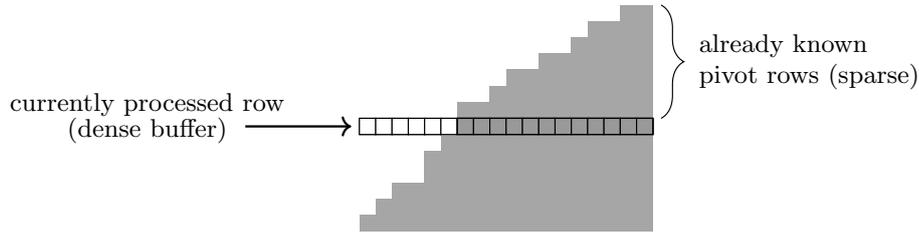

The forward elimination phase begins with a preprocessing step in which we permute the rows of $A$ such that,
in each row, the first nonzero entry appears directly below or to the left of the first nonzero entry in the row above. 
This reordering yields a matrix with an approximately triangular structure, as illustrated in Figure~\ref{fig:gauss}. 
We note that this preprocessing step is crucial for the efficiency of the algorithm.

We then process the reordered matrix row by row starting from the top.
To make row reductions more efficient, we maintain an auxiliary array \texttt{pivots} of length $n$ to keep track of already identified pivot elements.
In particular, \texttt{pivots[j]} stores the index of the row containing a pivot in column $j$, or $-1$ if no pivot has been assigned to that column yet.
This allows constant-time retrieval of pivot rows for any given column.
 
When processing a row $r$, we first identify the column index $j$ of its first nonzero entry. 
If column $j$ has no existing pivot, i.e, $\texttt{pivots[j]} = -1$, we do not perform any reductions on $r$. 
Instead, we update \texttt{pivots[j]} to reflect that row $r$ provides a pivot for column $j$, we normalize $r$ so that its pivot is $1$, and then proceed to the next row.
Otherwise, we fully reduce row $r$ using all previously identified pivots.

As also done in commutative F4 implementations~\cite{MP15,msolve}, we perform row reductions using a dense data structure.
In particular, we copy the current sparse row $r$ into a dense buffer -- an auxiliary array of length $n$, initialized to $0$. 
We then iterate over all positions of this buffer in ascending order, reducing all nonzero entries for which a pivot is known.

The buffer stores entries as 64-bit (signed) integers, which allows us to delay modular reductions and accumulate multiple arithmetic operations before 
applying expensive modulo operations.
To prevent overflows during this accumulation, we add a square of the field characteristic $p$ if a value becomes negative.
This is efficiently implemented via the branch-free update \texttt{buffer[j] += (buffer[j]  >>  63)  \& p*p}.

Once the buffer is fully reduced, we copy its content back into the sparse row $r$.
If the resulting row is nonzero, we normalize it to make the pivot equal to $1$ and update the \texttt{pivots} array accordingly.

\begin{remark}
While matrices arising in commutative F4 computations are already quite sparse, 
we observed that matrices in our setting are even sparser, with row densities rarely exceeding \SI{0.1}{\percent}. 
Due to this extreme sparsity, we experimented with removing the dense buffer entirely and performing all reductions directly within the sparse representation. 
However, this approach proved to be several orders of magnitude slower. 
We also tried hybrid strategies that tracked potential nonzero positions in the dense buffer to limit unnecessary work, but even those were outperformed by the fully dense approach. 
\end{remark}

Once all rows have been processed, the matrix is in row echelon form. 
To complete the computation of the reduced row echelon form, we first permute the rows again to restore the staircase shape illustrated in Figure~\ref{fig:gauss}, 
which may have been disrupted during the forward elimination phase. 
In this step, all zero rows are moved to the bottom.
Then we process the rows again from top to bottom, now reducing each row completely by all rows above it.
For this, we use the same dense reduction strategy as in the forward phase.

To take advantage of modern multi-core hardware, our implementation supports parallel Gaussian elimination, following the approach described in~\cite{MP15}.
Whenever a row is ready to be reduced, this reduction is launched as a separate task and added to a global work queue, which is processed in parallel by a thread pool.
Each thread is given its own local dense buffer to work in and all threads share access to a global \texttt{pivots} array.
After reducing a row, if the result is nonzero, we use an atomic \emph{compare-and-swap} (CAS) operation
(realized via C++'s \texttt{atomic\_compare\_exchange\_weak}) to update the global \texttt{pivots} array.
If this CAS operation fails, meaning that another thread has already registered a pivot at the same position, 
we reload the row into the dense buffer and reduce it further.

\paragraph{Mersenne Primes}

Although we delay modular reductions as much as possible, 
we observed that the modulo operation can still account for up to \SI{30}{\percent} of the total run time in matrix reduction.
To reduce this overhead, we exploit an arithmetic shortcut for a special class of primes.
Our implementation includes a dedicated optimization for \emph{Mersenne primes}, which are prime numbers of the form $p = 2^{b} -1$ for some positive integer $b$.
When such a prime is used as modulus, we replace the standard \texttt{\%} operator with a dedicated routine~\cite[Alg.~4]{ahle2021powerhashingmersenneprimes}.
For inputs $v < 2^{2b}$, the modular image $y = v \text{ mod }p$ can be computed as follows:
\begin{lstlisting}[xleftmargin=2em]
v1 = v + 1;
z  = ((v1 >> b) + v1) >> b;
y   = (v + z) & p;
\end{lstlisting}
Table~\ref{tab:mersenne} summarizes the benefits of this optimization on different CPUs and for different compilers. 
In our benchmark setting used in Section~\ref{sec:experiments} (AMD EPYC 7313 with GCC 13.3.0), it yields a speedup of $3.5\times$.
We have also tried to use~\cite[Alg.~5]{ahle2021powerhashingmersenneprimes} for modular reductions by \emph{pseudo-Mersenne primes} of the form
$2^{b} - c$ with $c > 1$ but this did not improve performance over the standard modulus operator.
% TODO: Only mention GCC

\begin{table}[tb]
  \centering
  \sisetup{round-mode=places,
    round-precision=2,
    round-pad = false
  }
  \begin{tabular}{ll|SSS}
    {CPU} & {Compiler} & {\;Standard [s]\;} & {Optimized [s]\; } & {Speedup} \\ 
    \midrule
% BEGIN RECEIVE ORGTBL mersenne
Intel i7-10700 & GCC 15.1.1 & 11.685667 & 1.311982 & 8.9068806\\
Intel i7-10700 & Clang 20.1.3 & 11.623475 & 1.375272 & 8.451767\\
Intel Xeon w5-2455X & GCC 13.3.0 & 4.709925 & 1.177109 & 4.0012650\\
AMD EPYC 7313 & Clang 18.1.3 & 4.057575 & 1.159551 & 3.4992639\\
AMD EPYC 7313 & GCC 13.3.0 & 4.057734 & 1.159754 & 3.4987885\\
Intel Xeon w5-2455X & Clang 18.1.3 & 4.708247 & 3.460993 & 1.3603746\\
MacBook Air M2 (ARM64)\;\; & AppleClang 16\;\; & 1.263310 & 1.491293 & 0.84712394\\
% END RECEIVE ORGTBL mersenne
  \end{tabular}
  \caption{Speedup of Mersenne prime optimization on different CPUs when reducing all integers in the range $2^{31} \pm 2^{16}$ modulo $p = 2^{31} - 1$.}
  \label{tab:mersenne}
\end{table}

\begin{comment}
#+ORGTBL: SEND mersenne orgtbl-to-latex :splice t :skip 2
| CPU                   | Compiler      | Standard [s] | Optimized [s] | Speedup [Standard/Optimized] |
|-----------------------+---------------+--------------+---------------+------------------------------|
| Intel i7-10700        | GCC 15.1.1    |    11.685667 |      1.311982 |                    8.9068806 |
| Intel i7-10700        | Clang 20.1.3  |    11.623475 |      1.375272 |                     8.451767 |
| Intel Xeon w5-2455X   | GCC 13.3.0    |     4.709925 |      1.177109 |                    4.0012650 |
| AMD EPYC 7313         | Clang 18.1.3  |     4.057575 |      1.159551 |                    3.4992639 |
| AMD EPYC 7313         | GCC 13.3.0    |     4.057734 |      1.159754 |                    3.4987885 |
| Intel Xeon w5-2455X   | Clang 18.1.3  |     4.708247 |      3.460993 |                    1.3603746 |
| MacBookAir M2 (ARM64) | AppleClang 16 |     1.263310 |      1.491293 |                   0.84712394 |
\end{comment}

\subsection{Multi-Modular Approach over the Rationals}

Gr\"obner basis computations over the rational numbers typically suffer from severe intermediate coefficient growth, which significantly impacts performance.
To address this issue, multi-modular techniques have been developed~\cite{Arn03,IPS11}.
These methods reduce computations over $\QQ$ to computations modulo different primes, thereby bounding the size of the coefficients, and subsequently reconstruct the rational results from their modular images.

In the commutative setting, multi-modular methods are typically used in an end-to-end fashion: a complete Gr\"obner basis is computed modulo different primes
and only the final rational Gr\"obner basis is reconstructed.
It has been shown~\cite{HV25} that this classical strategy fails in the noncommutative case because of the
infinite nature of Gr\"obner bases in free algebras.

Nevertheless, while an end-to-end multi-modular computation is not viable in our setting, the linear algebra within the F4 algorithm can still be performed in a multi-modular fashion. 
Our implementation uses the multi-modular algorithm presented in~\cite[Alg.~7.6]{Ste07} for computing the reduced row echelon form of rational matrices efficiently.
In this way, we can still circumvent expensive rational arithmetic and leverage fast modular computations.  
Moreover, we highlight that end-to-end methods lack a general and efficient test to verify the correctness of the reconstructed result 
and therefore yield correct results only with high probability~\cite{Arn03,msolve}.
In contrast, our method provides a simple verification criterion (see line~\ref{line:criterion} of Alg.~\ref{alg:rref-Q}) and is thus \emph{guaranteed} to return the correct result.

The multi-modular algorithm is summarized in Algorithm~\ref{alg:rref-Q}.
In this description, we denote by $H(A) = \max_{i,j} \lvert A_{i,j} \rvert $ the height of an integer matrix $A$.
For a correctness proof of  Algorithm~\ref{alg:rref-Q}, we refer to~\cite[Alg.~7.6]{Ste07}.
For further details on the reconstruction via Chinese remaindering and rational reconstruction, see, e.g.,~\cite[Ch.~5]{modernCA}.

\begin{algorithm}
\caption{Multi-modular Reduced Echelon Form}\label{alg:rref-Q}
\begin{algorithmic}[1]
\REQUIRE A matrix $A \in \QQ^{m \times n}$.
\ENSURE The reduced row echelon form of $A$.
\STATE $A' \leftarrow$ rescale $A$ to have integer entries\;
\STATE choose a finite set $P$ of primes\;
\WHILE{\textsc{True}} 
\FOR{$p \in P$}
	\STATE $R_{p} \leftarrow$ the reduced row echelon form of $A' \text{ mod } p$\label{line:ref}\;
\ENDFOR
\STATE discard any $R_{p}$ whose pivot column vector is not maximal
\footnotemark among all $R_{p}$, $p \in P$%
\STATE try to reconstruct from the modular echelon forms $R_{p}$, $p \in P$, a rational matrix~$R$ via Chinese remaindering and rational reconstruction\;
\IF {the reconstruction succeeds}
	\STATE $d \leftarrow$ the smallest positive integer such that $dR$ has only integer entries\;
		\IF{$H(dR) \cdot H(A') \cdot n < \prod_{p\in P} p$}\label{line:criterion}
	\RETURN $R$\;
	\ENDIF
\ENDIF
\STATE enlarge $P$ with primes not used so far\;
\ENDWHILE
\end{algorithmic}
\end{algorithm}

\paragraph{Tracer}
To optimize Algorithm~\ref{alg:rref-Q}, we have implemented a technique described in~\cite{msolve}:
During the very first modular computation in line~\ref{line:ref}, we construct a data structure called \texttt{tracer}, which records all rows of the matrix that reduce to zero.
In all subsequent modular echelon form computations, we then apply this \texttt{tracer} to immediately zero out the corresponding rows, avoiding any additional computations.
This optimization drastically speeds up the multi-modular algorithm, however the final output will only be correct with high probability. 
In particular, the result is guaranteed to be correct if the true reduced echelon form does not contain an entry whose denominator 
is divisible by the prime used to compute the \texttt{tracer}.
This happens with high probability and the number of bad primes is finite.
In our implementation, we use the Mersenne prime $p = 2^{31} - 1$ as the first prime for which the \texttt{tracer} is computed.
This allows us to exploit the fast Mersenne modular operation discussed at the end of Section~\ref{sec:gauss-Zp}.  
The tracer can be disabled to ensure correctness of the output.

\footnotetext{The pivot column vector of $R_{p}$ consists of the indices of all pivot columns of $R_{p}$ sorted in ascending order.
The maximality of such a vector is determined as follows: first the dimensions are compared (shorter vectors are smaller)
and vectors of the same dimension are ordered according to the reverse lexicographic order.}

\section{Proof Logging}
\label{sec:proofs}

Providing (efficient) certificates that a given set of polynomials forms a Gr\"obner basis remains an open problem. 
Aside from verifying the diamond lemma (Thm.~\ref{thm:diamond-lemma}) -- which essentially amounts to re-running the Gr\"obner basis algorithm -- no general method is currently known. 
We can, however, at least certify that each output polynomial computed by our implementation belongs to the ideal generated by the input using \emph{cofactor representations}.

\begin{definition}
A \emph{cofactor representation} of a polynomial $g \in K\<X>$ with respect to a set $F \subseteq K\<X>$ 
is an expression of the form
$g = \sum_{i=1}^k p_i f_i q_i$,
with $k \in \NN$, $p_i, q_i  \in K\<X>$, and $f_i \in F$.
\end{definition}
Such a representation serves as a certificate that $g \in (F)$.

Our library \name{} supports the computation of two kinds of cofactor representations for each output polynomial $g_{i}$:
\begin{enumerate*}[label=(\roman*)]
\item \label{item:cofactor-1} w.r.t.~the input $\{f_{1},\dots,f_{r}\}$ and all previously computed polynomials $\{g_{1},\dots,g_{i-1}\}$, or
\item  \label{item:cofactor-2} directly w.r.t.~the input $\{f_{1},\dots,f_{r}\}$.  
\end{enumerate*}

Both forms certify that the computed basis is indeed a subset of the input ideal (the latter directly, the former recursively).
These representations can be derived during the row echelon reduction step of the F4 algorithm (step~\ref{item:f4-reduction}), 
when the transformation matrix $T$ is computed such that $T \cdot A = \text{rref}(A)$. 
Type~\ref{item:cofactor-1} representations can be read off directly from the rows of $T$.
Type~\ref{item:cofactor-2} requires an additional backward substitution step.

\begin{remark}
The backward substitution step for computing representations of type~\ref{item:cofactor-2} can be computationally expensive and lead to an exponential increase in the size of the representations. 
Thus, representations of type~\ref{item:cofactor-1} are generally preferable.
%, as they incur less computational overhead.  
We nevertheless support both types as certain applications may require representations of type~\ref{item:cofactor-2}.
\end{remark}

\section{Experiments}
\label{sec:experiments}

We evaluate \name{} against the related tools from Section~\ref{sec:software}, 
focusing on \gbnp, \letterplace, \macaulay, \ncpoly, and \operatorgb.
We exclude \bergman and \opal from the evaluation as they are no longer maintained and were already outperformed by 
earlier versions of \letterplace~\cite{letterplace-old}.
We also omit \ncalgebra as initial experiments indicated that it is several orders of magnitude slower than the other tools 
(\ncalgebra required almost 3 hours to solve \texttt{braid3-11}, while \name{} needed \num{0.1} seconds).
Due to licensing restrictions, we could not evaluate \magma.
However, to offer a rough comparison, we include \magma timings from~\cite{letterplace}.

In Table~\ref{tbl:statistics}, we list the runtimes of each tool on the benchmarks from~\cite{letterplace}, all of which are part of the \symbolicdata project~\cite{symbolicdata}. 
For \name{}, we use the default configuration: single-threaded execution with the tracer for multi-modular computations and the Mersenne prime optimization enabled, and proof logging disabled.

Table~\ref{tbl:statistics-2} reports timings for the parallelized version of \name{}. 
For these tests, we use (larger) instances from the \symbolicdata project and we limit our comparison to \letterplace and \macaulay, as all other tools performed significantly worse on the smaller examples.
On these larger benchmarks, the single-threaded version of \name{} achieves speedups of up to $880\times$ over \letterplace and up to $510\times$ over \macaulay.
The parallelized version reaches speedups of up to $\num{1500}\times$ and $\num{920}\times$, respectively.
All reported times are real (wall\nobreakdash-clock) times.

All computations are performed over $\QQ$ w.r.t.~a degree-lexicographic monomial ordering.
To ensure termination, we impose a degree bound on the considered ambiguities in all computations.
The designated bounds are indicated by the number after the last ``\texttt{-}'' in the name of each example, 
e.g., in \texttt{braid3-11} we only consider ambiguities of degree~$\leq 11$. 

All experiments were run with a timeout of \SI{12}{\hour} (wall-clock
time) on a cluster of dual-socket AMD EPYC 7313 @ \SI{3.7}{\giga\hertz} machines
running Ubuntu 24.04 with a memory limit of \SI{30000}{\mega\byte} per
computation.

{\renewcommand{\arraystretch}{1.2}
\begin{table}[tb]
  \centering
    \sisetup{round-mode=places,
    round-precision=2,
    round-pad = false,
    detect-weight=true,
    detect-inline-weight=math
  }
  \scriptsize
  \begin{tabular}{l|SSSSSSS}
    % From Table selection2_f4ncgb_1

    % Get the benchmarks:
    % sqlite3 data.db "SELECT id,real,status,problem,logfile FROM selection2_f4ncgb_1 WHERE status = 'ok' ORDER BY problem" --markdown | grep (for f in cmp-with-magma/*.ms; set n (path basename $f); echo -n "$n\\|"; end; echo -n "NONONONO") -
    %
    % Max' Evil Macro to copy all of this from another buffer columnwise: 
    % (defalias 'datacopy
    %     (kmacro "C-w i v F y n C-w m p B n"))

    \toprule
     {Example}                & {\name{}}      & {\letterplace}  & {\macaulay}  & {\magma}  & {\gbnp} & {\ncpoly} & {\operatorgb}                        \\
 %                            & Time           & \%Ch            & \#             & \#{~}   & Time      & \# & Time & Guess & Eval & \%Co & m.It. \\
    \midrule
    \texttt{lascala\_neuh-10} & \bfseries 1.24  & 19.86            & 7.57  & 13.62         & 271.66  & 63.96      & 32.89                                  \\
    \texttt{serre-f4-15} 	  & 4.90           & \bfseries 3.66       & 5.63  & 8.96      & 2262.04 & 4303.90   & 8404.25                                 \\
    \texttt{serre-ha11-15} 	  & \bfseries 4.36 & 8.00                  & 12.05  & 5.80     & 2348.60 & 6258.89      & 7793.97                                 \\
    \texttt{serre-eha112-13}  & 1.51           & \bfseries 1.39          & 3.91 & 1.72    & 315.32  & 560.23      & 397.75                                  \\
    \texttt{4nilp5s-8} 		  & \bfseries 3.24 & 24.48                & 18.91 & 9.54       & 1126.18 & 548.48      & 147.17                                  \\
    \texttt{braidXY-11}		  & 15.86           & 87.97            & 1305.72   & \bfseries 4.20& 120.55  & 7649.53      & 5196.03                                 \\
    \texttt{ug2-x1x2x3x4-5}   & \bfseries 0.18 & 2.56                 & {TO}     & 0.83 & 2.61    & 26.33      & 6.53                                    \\
    \texttt{serre-e6-15}	  & 21.39          & 14.80 & \bfseries 14.51 & 40.99           & 2968.67 & 8862.33      & {TO}                                        \\
    \texttt{braid3-11} 		  & \bfseries 0.10 & 1.36           & 5.20   & 0.64           & 120.55  & 107.48      & 55.04                                   \\
    \texttt{ufn3-10} 		  & 2.60           & 34.76           & 6.05  & \bfseries 2.26  & 244.30  & 284.16      & 58.67                                   \\
    \texttt{ls3nilp-10} 	  & \bfseries 0.21 & 0.53               & 2.17  & 1.98        & 21.76   & 10.27      & 13.32                                   \\
\bottomrule
  \end{tabular}
  \caption{Timings (in sec) on benchmarks from~\cite{letterplace}. For \magma, we report the timings from~\cite{letterplace}. TO: $> \SI{43 200}{\second}$}
  \label{tbl:statistics}
  \end{table}
  
     {\renewcommand{\arraystretch}{1.2}
\begin{table}[tb]
    \sisetup{round-mode=places,
    round-precision=2,
    round-pad = false,
    detect-weight=true,
    detect-inline-weight=math
  }
  \centering
  \scriptsize
  % Table centering with stepping into margin: https://tex.stackexchange.com/a/672022
  \makebox[\linewidth][c]{%
  \begin{tabular}{l|SSSSSSS}
    % Get the benchmarks:
    % sqlite3 data.db "SELECT id,real,status,problem,logfile FROM selection2_f4ncgb_1 WHERE status = 'ok' ORDER BY problem" --markdown | grep (for f in hard-instances/*.ms; set n (path basename $f); echo -n "$n                                     \\|"; end; echo -n "NONONONO") -
    \toprule
    {\multirow{2}{*}{Example}}       &{\multirow{2}{*}{\letterplace}} &{\multirow{2}{*}{\macaulay}}   &  \multicolumn{5}{c}{\name{}} \\
    & & & {1 core} & {4 cores} & {8 cores} & {16 cores} & {32 cores}      \\
 %                                 & Time             & \#{~}                        & \%Ch                         & \#                           & Time                         & \#                            & Time & Guess & Eval & \%Co & m.It. \\
    \midrule
    \texttt{4nilp5s-10}         & 1282.44          & 875.09                        & 150.46                       & 79.47                        & 65.96                        & 63.03                         & \bfseries 57.44                    \\
    \texttt{braid3-16}             & 18953.19         & 14291.37                       & 105.46                        & 34.58                        & 23.96                        & 18.14               & \bfseries 17.69                              \\
    \texttt{braidX-18}             & {TO}           & {TO}                       & 1977.27                      & 601.12                       & 364.29                       & 260.65                         & \bfseries 239.30                   \\
    \texttt{braidXY-12}            & 1847.97          & 18887.61                        & 62.40                        & \bfseries 52.44              & 53.51                        & 52.98                         & 55.26                              \\
    \texttt{holt\_G3562h-17}       & {TO}          & {TO}                     & 25021.62                     & 12671.38                     & 8915.69                      & 6824.36                       & \bfseries 5715.45                  \\
    \texttt{lascala\_neuh-13} & 171.95           & 37.94                         & 9.84                         & 5.50                         & 4.98               & \bfseries 4.90                          & 5.25                                \\
    \texttt{lp1-15}                & 24166.85         & 33923.17                       & 266.44                       & 179.22                       & 161.98             & \bfseries 155.05                        & 168.91                             \\
    \texttt{lv2d10-100}            & {TO}          & 24930.25                        & 48.43                        & 27.68                        & \bfseries 26.98              & 47.65                         & 85.93                              \\
    \texttt{malle\_G12h-100}       & 4142.27          & 163.38                        & 89.93                         & 74.15              & 74.19                        & \bfseries 73.91                         & 79.86                              \\
    \texttt{serre-e6-17}           & 113.95 & \bfseries 84.99                       & 130.62                       & 130.53                       & 131.01                       & 128.97                        & 133.14                             \\
    \texttt{serre-ha11-17}         & 99.09            & 108.41                        & 38.18                        & 31.66                        & \bfseries 30.77                        & 30.84                & 31.87                              \\
\bottomrule
  \end{tabular}
  }
  \caption{Timings on larger benchmarks (in sec).  TO: $> \SI{43 200}{\second}$}
  \label{tbl:statistics-2}
  \end{table}  
 
 \section{Conclusion and Future Work}
 
We presented \name{}, a new open-source C++ library for Gr\"obner basis computations in free algebras
and discussed its implementation details and design choices.
Our experiments demonstrate that \name{} establishes a new state of the art for noncommutative Gr\"obner basis computations.
 
On some instances, however, \name{} seems to be rather memory inefficient. 
Reducing the memory footprint of the library is part of future work.
We also plan to integrate signature-based Gr\"obner basis algorithms, which have recently been generalized to the noncommutative setting~\cite{HV22,HV23}
and have proven useful in applications~\cite{HV24}.
Moreover, we want to extend the supported coefficient domains to include Euclidean domains, in particular, the integers.

 \section*{Acknowledgements}
 
We thank the anonymous reviewers for their feedback and comments.
We are also grateful to Yulia Mukhina for bringing the implementation in 
\macaulay{} to our attention. 

\bibliographystyle{splncs04}
\bibliography{references}
\end{document}